\documentstyle[sprocl,epsf]{article}

\def\Journal#1#2#3#4{{#1} {\bf #2}, #3 (#4)}

\def\NPB{{\em Nucl. Phys.} B}
\def\PLB{{\em Phys. Lett.}  B}
\def\PRL{\em Phys. Rev. Lett.}
\def\PRD{{\em Phys. Rev.} D}
\def\ZPC{{\em Z. Phys.} C}

\def\be{\begin{equation}}
\def\ee{\end{equation}}
\def\bea{\begin{eqnarray}}
\def\eea{\end{eqnarray}}

\begin{document}

\begin{center}
\noindent CERN-TH/98-399, December 1998, hep-ph/9906476\\[0.5cm]
\end{center}

\title{A TWO-LOOP APPLICATION OF THE THRESHOLD EXPANSION: THE 
BOTTOM QUARK MASS FROM $b\bar{b}$ PRODUCTION\footnote{
Talk presented at RADCOR98, Barcelona, September, 1998. 
The numerical result of this version 
supersedes the version printed in the proceedings volume, which was 
affected by a program error.}}

\author{M.~BENEKE, A.~SIGNER}

\address{Theory Division, CERN,\\ 
CH-1211 Geneva 23, Switzerland} 

\author{V.A.~SMIRNOV}

\address{Nuclear Physics Institute, Moscow State University,\\
119889 Moscow, Russia}


\maketitle\abstracts{We use the threshold expansion and 
non-relativistic effective theory to determine the bottom quark 
mass from moments of the $b\bar{b}$ production cross section 
at next-to-next-to-leading order in the (resummed) perturbative 
expansion, and including a summation of logarithms. For the 
$\overline{\rm MS}$ mass $\overline{m}_b$, we find 
$\overline{m}_b(\overline{m}_b)=(4.26\pm 0.12)\,$GeV.}

\section{Introduction}

In this talk we report on a determination of the bottom quark mass 
at next-to-next-to-leading order (NNLO) 
based on dispersion relations for $e^+ e^-\to b\bar{b}X$. This idea 
dates back to Ref.~\cite{NSVZ1} and relies on the relation 
\begin{equation}
\label{eq1}
\frac{12\pi^2}{n!}\,\frac{d^n}{d(q^2)^n}\,\Pi(q^2)_{\big| q^2=0} = 
\int_0^\infty \frac{ds}{s^{n+1}}\,R_{b\bar{b}}(s),
\end{equation}
between the two-point function of the bottom vector current and the 
inclusive bottom quark cross section mediated by a virtual photon, 
which is valid up to a small correction from $b\bar{b}$ 
radiation from light quarks. The left hand side can be computed in 
perturbation theory; the right hand side from data.

The parameters of the lowest $\Upsilon(nS)$ resonances are well-measured, 
but the continuum cross section above the threshold is not. Hence the 
experimental error of the right hand side decreases with increasing $n$, 
because higher moments weight lower $s$. When the integral over $s$ is 
saturated by the threshold region, the bottom quarks are non-relativistic, 
and $2 m_b/\sqrt{n}$ and $m_b/n$ appear as new momentum scales 
in the problem, related to the typical momentum $m_b v$ and 
non-relativistic energy $m_b v^2$ of the bottom quarks, respectively.
When $n$ is large enough such that $v\sim \alpha_s$, ordinary 
perturbation theory breaks down, because there are terms of the 
form $(\alpha_s/v)^k$ to all orders in perturbation theory. Even when 
$v$ is only small, but not as small as $\alpha_s$, one may want to resum 
those enhanced terms systematically.

Defining $v=(1-4 m_b^2/s)^{1/2}$, where $m_b$ is the $b$ quark pole mass, 
the cross section $e^+ e^-\to b\bar{b}X$ for $s\approx 4 m_b^2$ 
can be expanded in a double series in $\alpha_s$ and $v$:
\begin{equation}
\label{exp}
R_{b\bar{b}} = v \sum_{k=0}\,\sum_{l=-k} c_{kl}\,\alpha_s^k\,v^l
\times\,\mbox{logs of}\,\,v.
\end{equation}
In the resummed perturbative expansion, a N$^p$LO calculation has to account 
for all terms with $k+l\leq p$. Such a rearranged expansion has first 
been considered in this context in Ref.~\cite{VZ87} to LO accuracy and in 
Ref.~\cite{Vol95} to NLO accuracy. The step to NNLO accuracy implies 
new difficulties, because at this order the naive relativistic or 
non-relativistic approximations lead separately to divergent integrals. 
One has to be more precise about factorizing the two momentum regimes.

Powers of $v$ in (\ref{exp}) arise from ratios of momentum scales. 
We have to disentangle the contributions 
from the different scales in order to be sure that for a high-order 
loop graph, which cannot be calculated exactly, we have taken 
into account all terms with $k+l\leq 2$. In Ref.~\cite{BS98} we described 
how to expand a Feynman integral in $v$ to any given order, without 
calculating the exact result. The expansion procedure is based on 
decomposing the loop integral into contributions from different 
loop momentum regions and has a transparent interpretation in terms of 
effective theories. The present application, the perturbative calculation 
of $R_{b\bar{b}}$ in the threshold region, has much in common with a 
QED bound state calculation done with non-relativistic QED \cite{CL86,KN96}. 
The main difference is that we perform factorization in dimensional 
regularization, rather than using momentum space cut-offs and a photon 
mass as infrared regulator, and take advantage of the fact that 
scaleless integrals vanish in dimensional regularization. The absence 
of any explicit cut-off scale leads to a homogeneous expansion in $v$, 
manifest power counting and makes analytic calculations much easier at 
two loops. The unfamiliar dimensional renormalization of the Schr\"odinger 
problem is a negligible price to pay in comparison to these advantages, 
because it is not necessary to obtain the Green function exactly 
in $d$ dimensions.

In the following we summarize the calculation of $e^+ e^-\to b\bar{b}X$ 
at NNLO using these methods \cite{BSS99,BS99} and we determine 
the bottom quark mass by calculating the moment integral (\ref{eq1}). 
In addition to the NNLO terms in the sense of (\ref{exp}), we also 
sum all logarithms of $v$ of the form $(\alpha_s\ln v)^k$ and 
$\alpha_s\,(\alpha_s\ln v)^k$. A subset of NNLO logarithms of the form 
$\alpha_s^2\,(\alpha_s\ln v)^k$ can easily be summed as well, but 
a complete summation requires another difficult calculation to be done. 
This write-up 
concentrates on necessary formulae. A more general perspective on the 
problem, and a more complete list of references related to the 
subject can be found in Ref.~\cite{Ben98a}. While we were preparing this 
work, other NNLO calculations appeared \cite{PP98,Hoa98,MY98}. A brief 
comparison with these results is given at the end.

\section{Heavy quark production cross section near threshold}

\subsection{Classification of momentum regions}

The expansion (\ref{exp}) can be constructed by decomposing each 
Feynman integral that contributes to the vacuum polarization 
$\Pi(q^2)$ into contributions from the following momentum regions 
\cite{BS98}: 
\begin{eqnarray}
\label{terminology}
\mbox{{\em hard} (h):} && l_0\sim m_b,\,\,\vec{l}\sim m_b, \nonumber\\
\mbox{{\em soft} (s):} && l_0\sim m_b v,\,\,\vec{l}\sim m_b v,\nonumber\\
\mbox{{\em potential} (p):} && l_0\sim m_b v^2,\,\,\vec{l}\sim m_b v,\\ 
\mbox{{\em ultrasoft} (us):} &&  l_0\sim m_b v^2,\,\,\vec{l}\sim m_b v^2,
\nonumber
\end{eqnarray}
assuming a frame where $\vec{q}=0$. In dimensional regularization, 
the factorization of the different momentum regions is achieved by 
appropriate expansions of the Feynman integrands, without the need 
for explicit cut-offs, so that one can integrate over the whole 
integration domain in each region. As a consequence of this expansion, 
all terms have a definite homogeneous scaling behaviour in $v$. 

In the following, we integrate out first the hard, relativistic 
modes. This defines the NRQCD Lagrangian in dimensional regularization. 
In a second step, we also integrate out soft modes and potential 
gluon modes to arrive at another effective theory, `potential NRQCD' 
(PNRQCD), the QCD analogue of PNRQED, introduced in Ref.~\cite{PS}. 
The final result is obtained by computing the vacuum polarization in 
PNRQCD perturbation theory, with the Coulomb interaction treated 
non-perturbatively.

\subsection{Relativistic corrections}

At tree-level relativistic corrections arise from the anti-particle 
pole of the heavy quark propagator and the anti-particle components 
of the heavy quark four-spinor field. Beyond tree level, the hard loop 
momentum regions induce further terms in the non-relativistic effective 
theory. 

The heavy quark current correlation function $\Pi(q^2)$ contains 
1PI, hard subgraphs which (a) do not connect to the 
virtual photon vertex, (b) connect to one of the virtual photon vertices 
and (c) connect to both virtual photon vertices. The last possibility 
is of no interest, since it cannot contribute to the discontinuity 
of $\Pi(q^2)$ and hence is irrelevant for the computation of 
(\ref{eq1}).

Accounting for the hard subgraphs (a) leads to the familiar NRQCD 
Lagrangian \cite{TL,LMM}. At NNLO, the following terms are needed:
\begin{eqnarray}
\label{nrqcd}
{\cal L}_{\rm NRQCD} &=& \psi^\dagger \left(i D^0+\frac{\vec{D}^2}{2 m_b}
\right)\psi + \frac{1}{8 m_b^3}\,\psi^\dagger\vec{D}^4\psi
-\frac{d_1\,g_s}{2 m_b}\,\psi^\dagger\vec{\sigma}\cdot \vec{B}\psi
\nonumber\\
&&\hspace*{-1cm}
+\,\frac{d_2\,g_s}{8 m_b^2}\,\psi^\dagger\left(\vec{D}\cdot\vec{E}-
\vec{E}\cdot\vec{D}\right)\psi
+ \frac{d_3\,i g_s}{8 m_b^2}\,\psi^\dagger\vec{\sigma}\cdot\left(
\vec{D}\times\vec{E}-\vec{E}\times\vec{D}\right)\psi 
\nonumber\\[0.3cm]
&&\hspace*{-1cm}
+\,\,\mbox{antiquark terms}\,+ {\cal L}_{\rm light}.
\end{eqnarray}
Since we work in dimensional regularization, $\epsilon_{ijk}$ should 
be defined carefully. As long as we are only interested in $\Pi(q^2)$, 
we can avoid the problem by writing the Lagrangian in terms 
of anti-commutators of Pauli-matrices. Hence $\vec{\sigma}\cdot \vec{B}$ 
must be interpreted as $i/4\,[\sigma^i,\sigma^j]\,G^{ij}$. A  
similar interpretation holds for the spin-orbit interaction. The scaling 
of the interaction terms relative to $\psi^\dagger i D^0 \psi$ 
depends on whether the quark and gluons fields are considered as soft, 
potential or (in the case of gluons) ultrasoft. For potential quarks,
$\vec{D}^2/(2 m_b)$ is a leading order term, so its coefficient would 
be needed to order $\alpha_s^2$. However, since the coefficient is 1 
to all orders in PT, there is nothing to calculate. The chromomagnetic 
interaction is suppressed by only one power of $v$ for a soft gluon  
and by two powers of $v$ for an ultrasoft gluon. However, since 
one needs at least two chromomagnetic insertions to obtain a non-vanishing 
contribution to $\Pi(q^2)$, it is sufficient to know $d_1$ at tree level.
Since, for the time being, we sum logarithms of $v$ 
only at NLO and not at NNLO, we 
have $d_1=d_2=d_3=1.$

Accounting for the hard subgraphs (b) leads to the effective 
non-relativistic $\gamma^\star b\bar{b}$ coupling
\begin{eqnarray}
\label{current}
\bar{Q}\gamma^i Q &=& c_1\,\psi^\dagger \sigma^i\chi - 
\frac{c_2}{6 m_b^2}\,\psi^\dagger \sigma^i (i\vec{D})^2\chi + 
\ldots,
\end{eqnarray}
where the ellipsis refers to terms not needed for $\Pi(q^2)$ and at 
NNLO. At NNLO, we can use $c_2=1$, while $c_1$ is needed at 
order $\alpha_s^2$. The required matching calculation has been done in 
Refs.~\cite{BSS98,CM98}. The result is
\begin{equation}
\label{cc1}
c_1(\mu) = 1+ \left[c_1^{(1)}\frac{\alpha_s(m_b)}{4\pi} + 
\delta_1 \left(\frac{\alpha_s(m_b)}{4\pi}-
\frac{\alpha_s(\mu)}{4\pi}\right)\right] 
+c_1^{(2)}\left(\frac{\alpha_s(m_b)}{4\pi}\right)^{\!2},
\end{equation}
where 
\begin{eqnarray}
&& c_1^{(1)} = -\frac{32}{3}, \qquad \delta_1 = -\frac{1120\pi^2}{54 b_0},
\nonumber\\[0cm]
&& c_1^{(2)} = -\frac{712}{27}-\frac{2044\pi^2}{81}-\frac{224\pi^2}{9} 
\ln 2-\frac{2000\zeta(3)}{9}+\frac{176}{27}\,n_f 
\end{eqnarray}
and $b_0=11-2 n_f/3$. The term proportional to $\delta_1$ sums all 
next-to-leading logarithms of the form $\alpha_s (\alpha_s\ln v)^k$. There 
are no leading logarithms and this term 
is the only source of next-to-leading logarithms in the problem. 

\subsection{Instantaneous interactions}

The loop diagrams constructed from the NRQCD Lagrangian contain 
soft, potential and ultrasoft modes. We now integrate out the soft 
modes and potential gluons. As discussed in \cite{BS98,PS}, these 
modes give rise to instantaneous interactions. At NNLO, it is sufficient 
to match the (on-shell) quark-antiquark scattering amplitude to 
the required order. We do this order by order in $\alpha_s$, and, 
except for $a_2$ below, we 
have checked that the resulting potentials are the same whether we 
use Coulomb gauge or a general covariant gauge. The terms in the 
effective PNRQCD Lagrangian, which we need at NNLO, are given by
\begin{eqnarray}
\label{pnrqcd}
{\cal L}_{\rm PNRQCD} &=& \psi^\dagger \left(i \partial^0+
\frac{\vec{\partial}^2}{2 m_b}+
\frac{\vec{\partial}^4}{8 m_b^3}
\right)\psi + 
\chi^\dagger \left(i \partial^0-\frac{\vec{\partial}^2}{2 m_b}-
\frac{\vec{\partial}^4}{8 m_b^3}
\right)\chi 
\nonumber\\
&&\hspace*{-1.5cm}
+\int d^3\vec{r}\,\left[\psi^\dagger T^A\psi\right](\vec{r}\,)\,
\left(-\frac{\alpha_s}{r}\right)\left[\chi^\dagger T^A\chi\right](0) +
{\cal L}_{\rm light}^{\rm free} 
\\
&&\hspace*{-1.5cm}
+\int d^3\vec{r}\,\left[\psi^\dagger\psi\right]_{ij;ab}(\vec{r}\,)\,\,
\delta V_{ijkl,abcd}(r)\left[\chi^\dagger\chi\right]_{kl;cd}(0), 
\nonumber
\end{eqnarray}
where we indicated the spin and colour indices. In general, the PNRQCD 
Lagrangian contains local interactions of potential quarks and ultrasoft 
gluons, which give rise to retardation effects, but they contribute 
only at N$^3$LO, provided one counts $\alpha_s(m_b v^2)$ as $v$, as 
one does for $\alpha_s(m_b v)$ and $\alpha_s(m_b)$. Power counting 
confirms the well known fact that for $r\sim 1/(m_b v)$ and 
$\alpha_s\sim v$, the leading 
order Coulomb interaction is not suppressed relative to the free 
heavy quark Lagrangian. The instantaneous interaction in the third line 
can be treated perturbatively. At NNLO, we need all potentials of the form 
$\alpha_s^{2,3}/r$, $\alpha_s^2/r^2$ and $\alpha_s/r^3$, counting 
$\delta^{(3)}(r)$ as $1/r^3$. To compute $\Pi(q^2)$, only the 
colour singlet and spin-1 projection is needed. The result, 
obtained from matching 
the quark-antiquark scattering amplitude, is, in momentum space, 
\begin{eqnarray}
\delta\tilde{V}_{ijkl,abcd}(\vec{p},\vec{q}) &=& 
\delta_{ij}\delta_{kl}\delta_{ab}\delta_{cd} \cdot 
\frac{16\pi\alpha_s}{-3\vec{q}^{\,2}} \cdot
\nonumber\\
&&\hspace*{-2cm}
\Bigg[\frac{\alpha_s}{4\pi}\left(a_1-b_0\ln\frac{\vec{q}^{\,2}}{\mu^2}
\right) + 
\left(\frac{\alpha_s}{4\pi}\right)^{\!2}\bigg(a_2-(2 a_1 b_0+b_1) 
\ln\frac{\vec{q}^{\,2}}{\mu^2}+
b_0^2 \ln^2\frac{\vec{q}^{\,2}}{\mu^2}
\bigg)
\nonumber\\
&&\hspace*{-2cm}
+\,\frac{\alpha_s}{4\pi}\,\frac{\pi^2 \,|\vec{q}\,|}{m_b}\,
\left(\frac{\vec{q}^{\,2}}{\mu^2}\right)^{-\epsilon}\left(
\frac{7}{3}+\epsilon \left(-\frac{5}{3}+\frac{14}{3}\ln 2\right)\right)
+\frac{\vec{p}^{\,2}}{m_b^2}
\\
&&\hspace*{-2cm}
-\,\frac{\vec{q}^{\,2}}{m_b^2} 
\left(d_1^2\left(\frac{1}{6}+\frac{5\epsilon}{18}\right)+
\frac{1}{4} (1+d_2)\right)
\Bigg],
\nonumber
\end{eqnarray}
where $b_1=102-38 n_f/3$ and $a_1$ and $a_2$ denote the one-loop and 
two-loop \cite{Sch98} radiative corrections to the Coulomb potential,  
respectively. The potentials 
more singular than $1/r$ lead to ultraviolet 
divergent integrals in PNRQCD 
perturbation theory. These divergences cancel with divergences 
that arise in the calculation of $c_1$ at the 
two-loop order. To obtain the correct finite terms, it is necessary to 
compute the potentials that lead to divergent insertions to order 
$\epsilon$, where $d=4-2\epsilon$ is the number of space-time 
dimensions.

\subsection{Perturbation theory with potential NRQCD}

The heavy quark current correlation function is now obtained from the 
two-point functions of the effective currents (\ref{current}), computed 
with the PNRQCD Lagrangian. Because the unperturbed PNRQCD 
Lagrangian contains the Coulomb interaction, the 
propagator for a $b\bar{b}$ pair is the 
Coulomb Green function $G_c(\vec{x},\vec{y};E)$, where 
$E=\sqrt{q^2}-2 m_b$. At LO in PNRQCD perturbation theory, one finds 
\begin{equation}
\Pi(q^2) = \frac{3}{2 m_b^2}\,G_c(0,0;E).
\end{equation}
The Green function at the origin is ultraviolet divergent. We need 
its value in dimensional regularization, with $\overline{\rm MS}$ 
subtractions. To this end we write all integrals first in momentum space 
and note that all divergences can be removed by counterterms that 
involve only a finite number of loops. In the case of $G_c(0,0;E)$, 
for example, all diagrams with two or more gluons exchanged  
are finite. The result in the $\overline{\rm MS}$ scheme is
\begin{equation}
\label{g0}
G_c(0,0;E) = -\frac{m_b^2 \alpha_s}{3\pi}\left[
\frac{1}{2\lambda}+\frac{1}{2}\,\ln\frac{-4 m_b E}{\mu^2}-\frac{1}{2}+
\gamma_E+\psi(1-\lambda)\right],
\end{equation}
where $\lambda=2\alpha_s/(3\sqrt{-E/m_b})$. The constant $-1/2$ is 
scheme-dependent. Since the cross section requires only the 
discontinuity of $\Pi$, the subtraction procedure is irrelevant for 
the moments. However, essentially the same procedure can be used 
to obtain more complicated integrals in dimensional regularization, 
which are needed at NNLO. In particular, the pole parts of 
the divergent potential insertions are all proportional to 
$G_c(0,0;E)$ as they have to be in order that the $\mu$ dependence 
that comes from (\ref{cc1}) cancels at NNLO.

To obtain the final result at NNLO one has to compute: (i) a single 
insertion of the kinetic energy correction $\vec{\partial}^4/(8 m_b^3)$ to 
the quark propagator; (b) the single and double insertion of the 
terms in $\delta V(\vec{p},\vec{q})$, which are suppressed by one 
power of $v\sim \alpha_s$ compared to the LO Coulomb potential; 
(c) the single insertion of all terms in $\delta V(\vec{p},\vec{q})$, 
which are suppressed by two powers of $v\sim \alpha_s$. The result is 
an expression for $R_{b\bar{b}}$ that can be cast into the form
\begin{equation}
\label{rr}
R_{b\bar{b}} = \alpha_s \left\{f_0(\lambda,l) + 
\alpha_s \,f_1(\lambda,l)+ 
\alpha_s^2 \,f_2(\lambda,l)\right\},
\end{equation}
where $l=\ln(-4 m_b E/\mu^2)$. One can then compute the integral 
(\ref{eq1}) and compare it with its experimental value. Note that 
$R_{b\bar{b}}$ contains a continuum starting at $4 m_b^2$ and 
an infinite series of $b\bar{b}$ bound state poles, which are 
included in the integral. In (\ref{rr}) the denominators 
of the bound state pole contribution are expanded around the LO 
pole position. Before we turn to a numerical analysis of the 
moments of (\ref{rr}), we have to discuss the issue of mass 
renormalization.

We also mention the following checks we performed on (\ref{rr}). 
We re-expanded (\ref{rr}) to order $\alpha_s^2$ and found agreement 
with the cross section near threshold computed to this order in 
Ref.~\cite{CM98}. We also obtain the bound state pole position and 
its residue in analytic form and confirm the result of 
Ref.~\cite{MY98}. Both together are strong checks that the 
factorization in dimensional regularization has been done correctly. 
Conversely, our calculation provides an independent check of 
the result of Ref.~\cite{CM98}, which has been used in the other 
NNLO calculations \cite{PP98,Hoa98,MY98}, which followed the  
`direct matching' procedure suggested in Ref.~\cite{HT98}, rather 
than factorization in dimensional regularization.

\section{Potential subtracted quark mass}

The heavy quark production cross 
section near threshold is conventionally 
expressed in terms of the quark pole mass. If one uses another mass 
renormalization convention that differs from the pole mass by $\delta m$, 
one finds terms of the form $(\delta m/E)^k$, which modify the 
structure of (\ref{exp}), and which seem to complicate the resummation. 
However, the pole mass is known \cite{BB94,BSUV94} to be more 
infrared sensitive than the heavy quark production cross section itself 
and one therefore expects that a badly convergent series expansion 
for $R_{b\bar{b}}$, when expressed in terms of $m_b$, would prevent us 
from extracting $m_b$ accurately. The convergence should be improved, 
when $R_{b\bar{b}}$ is expressed in terms of a less IR sensitive 
mass parameter. To implement this observation in the calculation, we 
make use of a systematic cancellation of infrared contributions to the 
pole mass and the Coulomb potential in coordinate space \cite{B98,HSSW98} 
and define the potential-subtracted (PS) quark mass $m_{b,\rm PS}(\mu_f)$ 
by \cite{B98}
\begin{eqnarray}
\label{mass}
m_b &=& m_{b,\rm PS}(\mu_f) -\frac{1}{2}\int\limits_{|\vec{q}\,|<\mu_f} 
\!\!\!\frac{d^3\vec{q}}{(2\pi)^3}\,\tilde{V}(q) 
\nonumber\\
&=& m_{b,\rm PS}(\mu_f) + \frac{4\alpha_s}{3\pi}\,\mu_f 
\left[1+\frac{\alpha_s}{4\pi}\,\delta m_1 + 
\left(\frac{\alpha_s}{4\pi}\right)^{\!2}\,\delta m_2 +\ldots\right].
\end{eqnarray}
Explicit expressions for $\delta m_{1,2}$ can be found in Ref.~\cite{B98}. 
Note that $m_b-m_{b,\rm PS}(\mu_f)$ is proportional to a subtraction 
scale $\mu_f$, which must be chosen to be smaller than $m_b v$. We insert 
(\ref{mass}) into (\ref{rr}) and expand the small correction terms 
involving $\delta m_{1,2}$. However, the term 
$\Delta=4\alpha_s\mu_f/(3\pi)$ is 
not expanded, when $m_b$ is replaced in $E$, $\lambda$ or $l$, because 
$\Delta$, which counts as order $m_b v^2$, is of the same order as 
$E=\sqrt{q^2}-2 m_b$. The result is an expression of the same form 
as (\ref{rr}), but with $m_{b,\rm PS}(\mu_f)$ as input parameter. If our 
expectation is correct, the expansion (\ref{rr}) should be more 
convergent in this new variable. We then extract $m_{b,\rm PS}(\mu_f)$ 
from comparison with the data, and finally use the known 2-loop 
relation between the pole mass and the $\overline{\rm MS}$ mass 
\cite{GBGS90} to convert to the $\overline{\rm MS}$ mass
$\overline{m}_b(\overline{m}_b)$.

\section{Results and discussion}

There are restrictions on the value of $n$ that can be chosen for 
the moments (\ref{eq1}). Comparison of the NNLO threshold approximation 
of the cross section up to order $\alpha_s^2$ with the `exact' result 
\cite{CKS97} shows that the threshold approximation provides a reasonable 
approximation as long as one is less than 2 GeV away from threshold. 
This restricts $n\geq 6$, conservatively. Requiring that all scales of 
the problem are larger than $\Lambda_{QCD}$ restricts $n\leq 10$. 

The experimental moments are obtained from the known masses and decay 
constants of the $\Upsilon$ resonances below the open $b\bar{b}$ 
threshold. The continuum cross section is parametrized by the 
constant value $0.4\pm 0.2$. (The asymptotic value for $E\to \infty$ 
is $1/3$.) For $n=8$ the continuum cross section contributes only 
about 10\% to the experimental moment integral.

\begin{figure}[p]
\caption[dummy]{\label{fig1} 
The 8th moment (in GeV$^{-16}$) as function of the bottom quark 
pole mass (in GeV) (upper figure) and the bottom quark PS mass 
(lower figure) at $\mu_f=2\,$GeV. The solid 
curve is for the scale $\hat{\mu}=2 m_b/\sqrt{8}$, the outer (dash-dotted) 
curves show the result, when the scale is varied by a factor of 2 
in both directions. The inner (dotted) curves correspond to $2\hat{\mu}/3$ and 
$3\hat{\mu}/2$. The experimental moment is given by the grey bar.}
   \vspace{-3.5cm}
   \epsfysize=12cm
   \epsfxsize=8cm
   \centerline{\epsffile{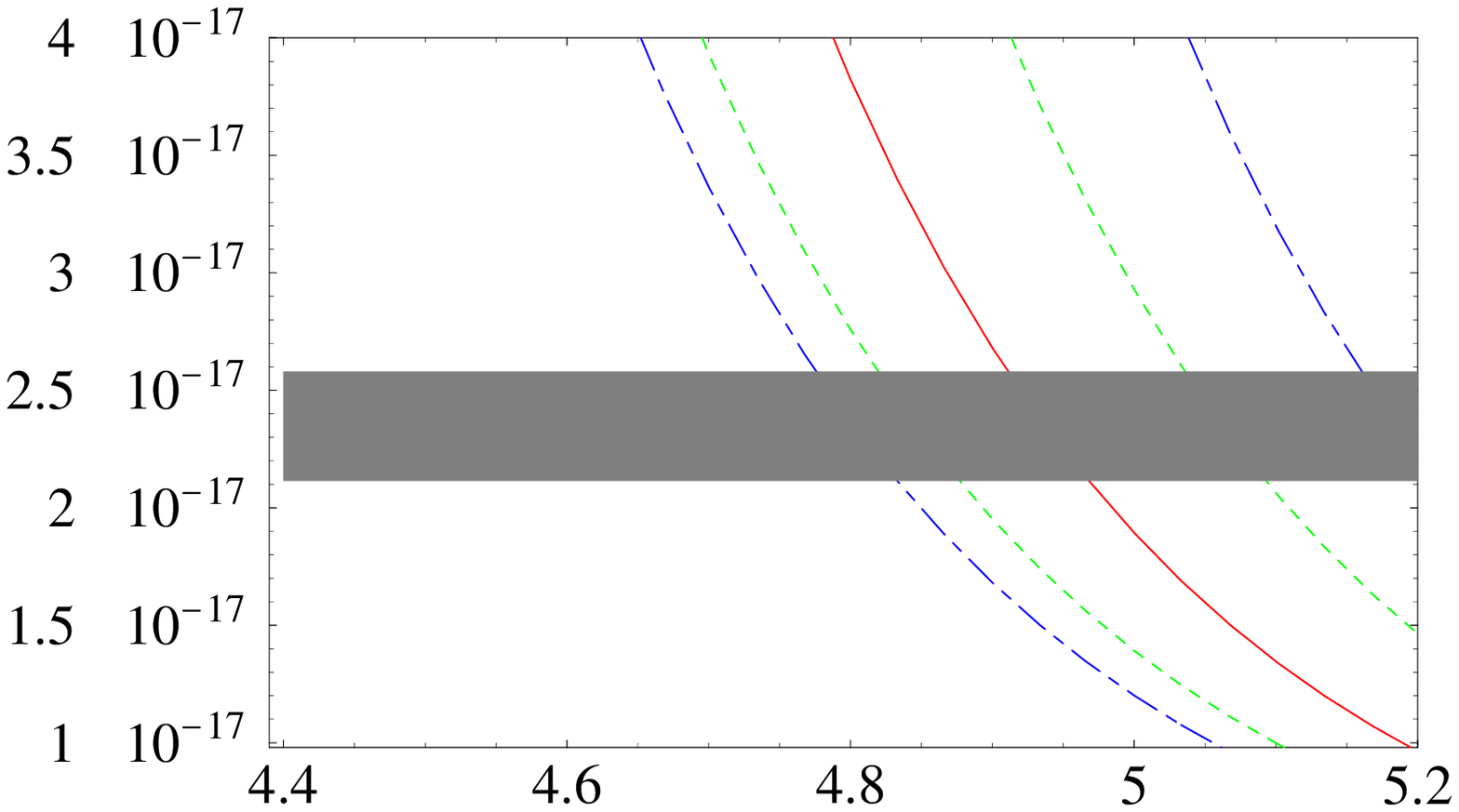}}
   \vspace*{-7cm}
   \epsfysize=12cm
   \epsfxsize=8cm
   \centerline{\epsffile{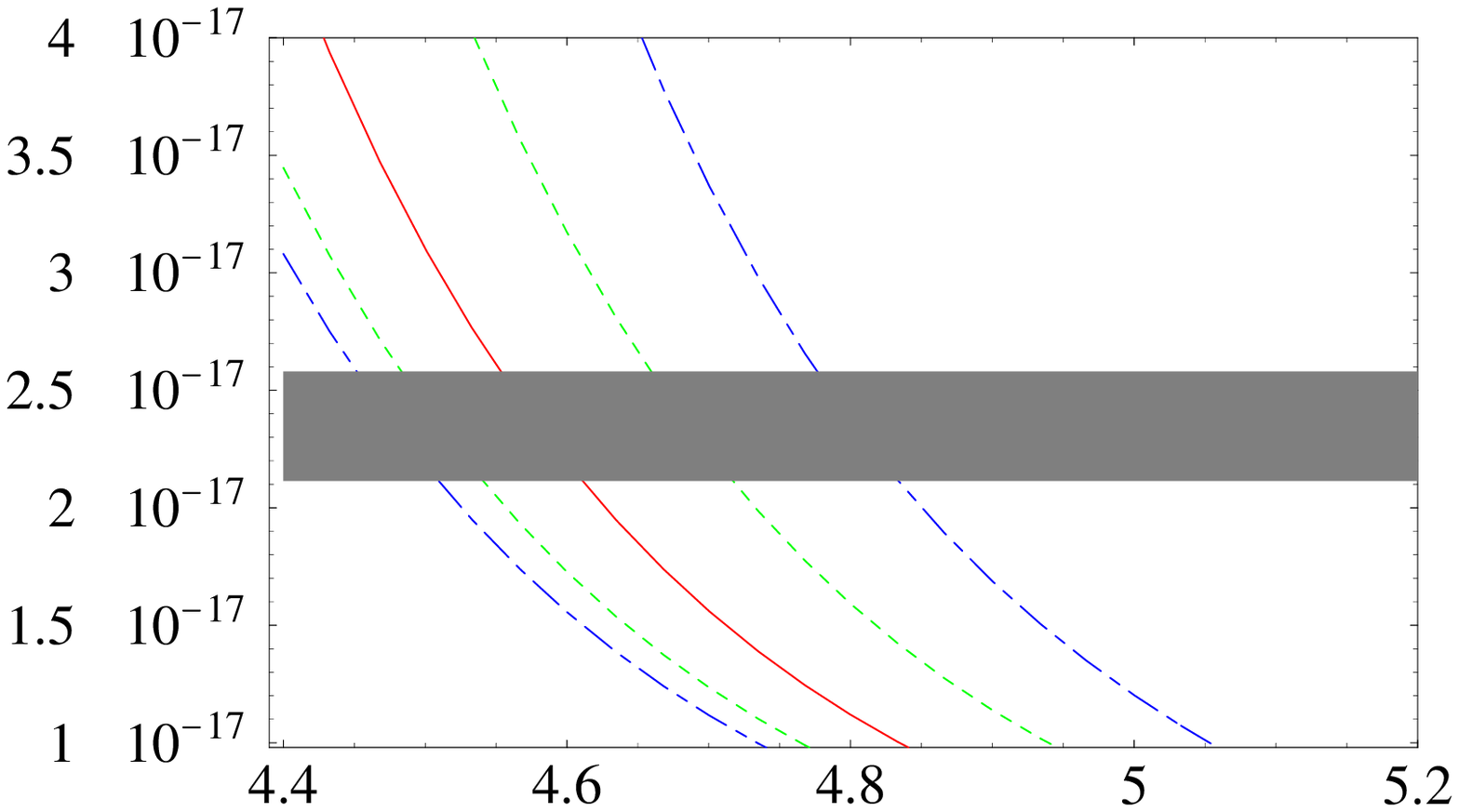}}
   \vspace*{-3.5cm}
\end{figure}

In Fig.~\ref{fig1}, we show the result for the 8th moment as a function 
of the quark mass. In the upper figure, the result is  plotted as a 
function of the pole mass $m_b$, in the lower figure as a function of 
the PS mass $m_{b,\rm PS}(2\,\mbox{GeV})$. For a given moment $n$, the 
integral (\ref{eq1}) implies the characteristic energy scale $E\sim m_b/n$. 
The logarithms that appear in the PNRQCD integrals then suggest that 
the scale $\mu$ is chosen as $2 m_b/\sqrt{n}$. This choice is 
shown as the solid line in the figure, together with variations 
by a factor $1/2,2/3,3/2,2$. In case of the pole mass as an input 
parameter, the scale dependence is very large. Including a (small in 
comparison) error due to the variation of $\alpha_s(M_Z)=0.118\pm 0.003$, 
we determine $m_b=(4.97\pm 0.17)\,$GeV, where the scale error is 
estimated from a variation between $2\,$GeV and $2\hat{\mu}$. 
The scale dependence is 
somewhat reduced, but still large, with the PS mass as input. Our result is
\begin{equation}
m_{b,\rm PS}(2\,\mbox{GeV}) = (4.60 \pm 0.125 (\mbox{scale}) \pm 
0.03 (\alpha_s) \pm 0.03 (\mbox{exp.}))\,\mbox{GeV},
\end{equation}
which translates into 
\begin{equation}
\label{result}
\overline{m}_b(\overline{m}_b)=(4.26\pm 0.12)\,\mbox{GeV}
\end{equation}
for the $\overline{\rm MS}$ mass. We have included an estimate of 
$(40\pm 40)\,$MeV for the unknown 3- and 4-loop terms in the relation 
between the PS and the $\overline{\rm MS}$ mass, which are needed here, 
because the sum rule determines the PS mass with a parametric accuracy of 
order $m_b\alpha_s^4$.  The central value is stable against 
variations of the order of the moments. The large remaining scale dependence 
can be traced to the scale dependence of the theoretical prediction 
of the first bound state residue. This is discussed in more detail in 
Ref.~\cite{BS99}.

\begin{table}[t]
\addtolength{\arraycolsep}{0.2cm}
\renewcommand{\arraystretch}{1.7}
$$
\begin{array}{|c|c|c|c|}
\hline & m_b & \overline{m}_b(\overline{m}_b) & \mbox{Remarks}\\ 
\hline
\mbox{This work} & 4.97\pm 0.17 & 4.26\pm 0.12 & \mbox{Sum rules}  \\
\mbox{PP98 \cite{PP98}} & 4.80\pm 0.06 & \mbox{--} & \mbox{Sum rules} \\
\mbox{H98 \cite{Hoa98}} & 4.88\pm 0.10 & 4.25\pm 0.09 & \mbox{Sum rules}  \\
\mbox{MY98 \cite{MY98}} & \mbox{--}  & 4.20\pm 0.10 & \mbox{Sum rules}  \\
\hline
\mbox{JP97/98 \cite{JP95,JP98}} & 4.60\pm 0.02 & 4.19\pm 0.06  &  
\mbox{Sum rules, no resummation}\\
\hline
\mbox{PY98 \cite{PY98}} & 5.00^{+0.10}_{-0.07} & 4.44^{+0.03}_{-0.04} & 
\Upsilon(\mbox{1S})\mbox{ mass} \\
\mbox{MS98 \cite{MS98}} & \mbox{--}  & 4.41\pm 0.11  & 
\mbox{lattice HQET} \\
\hline
\end{array}
$$
\caption[dummy]{\label{tab1}\small 
Bottom quark mass values (in GeV) obtained from NNLO calculations.}
\end{table}

In Tab.~\ref{tab1} we compare our result to other bottom quark mass 
results, choosing only NNLO calculations (although NNLO may not 
imply a NNLO resummation in some cases). The first four entries all 
refer to the sum rule method with NNLO resummation. There are several 
differences between the present and previous implementations of the 
NNLO moments. The summation of 
NLO logarithms of $v$, which has not been done in 
Refs.~\cite{PP98,Hoa98,MY98}, turns out to be a small effect; leaving it 
out would increase our result by up to $10\,$MeV. 
Refs.~\cite{PP98,Hoa98} use the `old' value of $a_2$ \cite{Pet97}, 
which is most likely incorrect \cite{Sch98}. Again, the corrected 
value for $a_2$ shifts the extracted quark mass downwards by only 
15 MeV. More significant differences arise in the treatment of the 
short-distance factor and the bound state pole $\delta$-functions. 
For a detailed discussion we refer to Ref.~\cite{BS99}.

The sum rule calculation of Refs.~\cite{JP95,JP98} does not include 
a non-relativistic resummation. It leaves out, in particular, the 
contribution from the Coulomb poles. This seems to be the main reason 
why the bottom masses of JP97/98 come out small. The large moments 
used in Refs.~\cite{JP95,JP98} are completely dominated by the 
Coulomb pole contribution to (\ref{rr}). For the 8th moment, we 
find that our pole quark mass would decrease by roughly 400 MeV, if we 
left out the Coulomb pole contribution.

The last two entries in the table come from other methods. 
Ref.~\cite{PY98} uses the NNLO result for the Coulomb 
pole position and compares it with the $\Upsilon(\mbox{1S})$ mass. 
The error quoted in Ref.~\cite{PY98} is probably underestimated, because 
the scale $\mu^2$ is allowed to vary only by $\pm 25\%$ rather than a factor 
of 2. The origin of the large value of the $\overline{\rm MS}$ quark 
mass in Ref.~\cite{PY98} will be explained in Ref.~\cite{BS99}.  
The last entry refers to the first complete NNLO quark mass 
determination from the lattice, based on the $B$ meson mass and the 
lattice measurement of the static energy of a heavy quark. 

\section*{Acknowledgments}

This work was supported in part by the EU Fourth Framework 
Programme `Training
and Mobility of Researchers', Network `Quantum Chromodynamics and the Deep
Structure of Elementary Particles', contract FMRX-CT98-0194 (DG 12 - MIHT).

\end{document}